\documentclass[12pt,a4,sans]{article}
\usepackage[latin1]{inputenc}
\usepackage{amssymb}
\usepackage{amsmath}
\usepackage{amsthm}
\usepackage{amsfonts}
\usepackage[pdftex]{graphicx}
\usepackage{geometry}
\usepackage{bbold}
\usepackage{braket}
\usepackage{enumerate}
\usepackage[usenames,dvipsnames]{xcolor}
\usepackage{setspace}
\usepackage[backgroundcolor=white,linecolor=black]{todonotes}
\usepackage[in]{fullpage}
\usepackage{hyperref}
\usepackage{array}
\usepackage{cancel}
\usepackage{wrapfig}
\usepackage[font=small,labelfont=bf]{caption}
\usepackage{anysize}
\usepackage{color}

\numberwithin{equation}{section}

\marginsize{2.4cm}{2.4cm}{2cm}{2cm}

\hypersetup{
colorlinks=true,
linktoc=page,
citecolor=Blue,
linkcolor=Blue,
urlcolor=Blue} 

\makeatletter 
\renewcommand{\maketitle} 
 { \begingroup \begin{center} \large {\bf \@title}
 	\vskip 5pt \large \@author \\ \vskip 5pt \@date \end{center}
   \vskip 5pt \endgroup \setcounter{footnote}{0} }
\makeatother 

\newcommand{\comments}[1]{}
\newcommand{\la}{\langle}
\newcommand{\ra}{\rangle}

\newcommand{\Tr}{\text{Tr}}

\newcommand{\be}{\begin{equation}}
\newcommand{\ee}{\end{equation}}

\def\beqa{\begin{eqnarray}}
\def\eeqa{\end{eqnarray}}
\def\beq{\begin{equation}}
\def\eeq{\end{equation}}

\def\Tr{{\rm Tr}}
\def\one{\mbox{1 \kern-.59em {\rm l}}}
 
\newcommand{\bea}{\begin{eqnarray}}
\newcommand{\eea}{\end{eqnarray}}

 \def\cN{{\cal N}}

\def\uno{\mbox{1 \kern-.59em {\rm l}}}

\def\one{1\!\!1\,\,}

\def\bcomment#1{}

\usepackage{slashed}
\usepackage{caption}

\usepackage[nottoc,notlot,notlof]{tocbibind}
\usepackage[nosort]{cite}
\usepackage{color}
\usepackage{bbm}
\usepackage{parskip}
\usepackage{mathtools}

\graphicspath{{./Images/}}

\long\def\symbolfootnote[#1]#2{\begingroup%
\def\thefootnote{\fnsymbol{footnote}}\footnote[#1]{#2}\endgroup}

\begin{document}

\begin{flushright}
QMUL-PH-15-11\\
DCPT-15/41
\end{flushright}

\vspace{20pt}

\begin{center}

{\Large \bf  On Yangian symmetry  of scattering amplitudes and    }\\
\vspace{0.3 cm}
{\Large \bf  the dilatation operator in  $\cN=4$ super Yang-Mills}

\vspace{45pt}

{\mbox {\bf  Andreas Brandhuber$^{a}$,  Paul Heslop$^{b}$, Gabriele Travaglini$^{a,b}$ and  Donovan Young$^{a}$}}%
\symbolfootnote[4]{
{\tt  \{ \tt \!\!\!a.brandhuber, g.travaglini, d.young\}@qmul.ac.uk, paul.heslop@durham.ac.uk}
}

\vspace{0.5cm}

\begin{quote}
{\small \em
\begin{itemize}
\item[\ \ \ \ \ \ $^a$]
\begin{flushleft}
Centre for Research in String Theory\\
School of Physics and Astronomy\\
Queen Mary University of London\\
Mile End Road, London E1 4NS, United Kingdom
\end{flushleft}
\item[\ \ \ \ \ \ $^b$]
Department of Mathematical Sciences\\
Durham University\\
South Road, Durham DH1 3LE, United Kingdom

\end{itemize}
}
\end{quote}


\vspace{40pt}

{\bf Abstract}
\end{center}

\vspace{0.3cm} 

\noindent

\noindent
It is known that the  Yangian of $PSU(2,2|4)$ is a symmetry of the tree-level $S$-matrix  of $\mathcal{N}=4$ super Yang-Mills. On the other hand, the complete one-loop dilatation operator in the same theory  commutes with the level-one Yangian generators only up to certain boundary terms found  by Dolan, Nappi and Witten. Using a result by Zwiebel, we show how the Yangian symmetry of the tree-level $S$-matrix of $\mathcal{N}=4$ super Yang-Mills  implies precisely the Yangian invariance, up to boundary terms,  of the one-loop dilatation operator.

\setcounter{page}{0}
\thispagestyle{empty}
\newpage



%


\section{Introduction}

The study of ${\cal N}=4$ supersymmetric Yang-Mills (SYM) theory  has been
dominated by two broad strands of research -- the first concentrating on the
anomalous dimensions of local operators (i.e.~the spectral problem)
and their correlation functions, and the second investigating the
scattering amplitudes of the theory. The
successes in these two areas have been considerable in their own
right, and at the current time there is  vigorous  activity  focussing on making
connections between them in order to deepen our understanding of this
fascinating quantum field theory.

In the planar limit the spectral problem is believed to be
integrable. This was first shown at one loop in \cite{Minahan:2002ve} for a particular sector of the theory. The complete one-loop dilatation operator was later computed in \cite{Beisert:2003jj},  following earlier results in \cite{Beisert:2003tq}, and later shown in \cite{Beisert:2003yb} to describe a $PSU(2,2 | 4)$ super spin chain. 
The one-loop dilatation operator is invariant under the (free) superconformal symmetry, and in fact this condition puts strong constraints on  its form. 

One of the key features of integrability is the existence of an
infinite hierarchy of non-local charges $Q^A$ built upon the basic local (or
level-zero) $PSU(2,2|4)$ Noether charges  $J^A$  of the theory. These non-local
charges, together with the local ones, obey a 
Yangian algebra which in the context of the one-loop  dilatation operator  $H$ was described 
in \cite{Dolan:2003uh}. Interestingly, it was found in that  paper that $H$ commutes with these additional  non-local charges   up to certain boundary terms,   
\beq
\label{main}
[Q^A, H]  \sim J^A_1 - J^A_L
\ , 
\eeq
where $L$ denotes the length of the chain (or number of fields in the operator). One intriguing aspect of this relation, which we will return to later, is that it mixes tree-level and one-loop quantities  \cite{Dolan:2004ys}.

The study of scattering amplitudes in ${\cal N}=4$ SYM started off
independently from considerations of  integrability, but has
recently begun to be connected to it in various ways. 
An important discovery was that of dual superconformal symmetry of the $\cN=4$ SYM $S$-matrix. This was 
conjectured in \cite{Drummond:2008vq} and tested in several cases, and shortly after proved at tree level in  \cite{Brandhuber:2008pf}. At one loop the symmetry is broken because of the presence of infrared divergences in the amplitudes, and the breaking is controlled by a dual conformal Ward identity proposed in \cite{Drummond:2007au} and confirmed with a direct amplitude calculation at one loop in \cite{Brandhuber:2009kh}.
Importantly,  in \cite{Drummond:2009fd}  the  standard and dual superconformal symmetries were embedded into the  Yangian  of $PSU(2,2|4)$. Explicit expressions of the level-one  generators were constructed and shown to be related to the generators of the dual superconformal algebra. At tree level the symmetry is slightly broken 
\cite{Bargheer:2009qu} due to collinear singularities of the amplitudes, leading to anomalies that are supported only on special kinematic configurations. As mentioned earlier, at one loop infrared divergences lead to additional anomalies. Interestingly, these violations can be absorbed into appropriate redefinitions of the Yangian generators both at tree level \cite{Bargheer:2009qu} and one loop \cite{Beisert:2010gn}.

The presence of a Yangian symmetry  on the dilatation operator and the amplitude sides makes one naturally think that these  symmetries are the manifestation of a single underlying Yangian symmetry of the theory. However these two symmetries are seemingly realised in a different manner, given \eqref{main} and the fact that on the amplitude side, the symmetry can be realised exactly, with the Yangian generators annihilating the amplitudes (divided by the MHV part). The goal of this paper is that of reconciling these two situations by finding a proof of \eqref{main} which relies on the Yangian symmetry of the tree-level $S$-matrix of $\cN=4$ SYM, therefore substantiating  the connection between the Yangians of the spin chain and the amplitudes. 

\begin{figure}
\begin{center}
\includegraphics[bb=50 70 540 355, height=1.5in, clip=true]{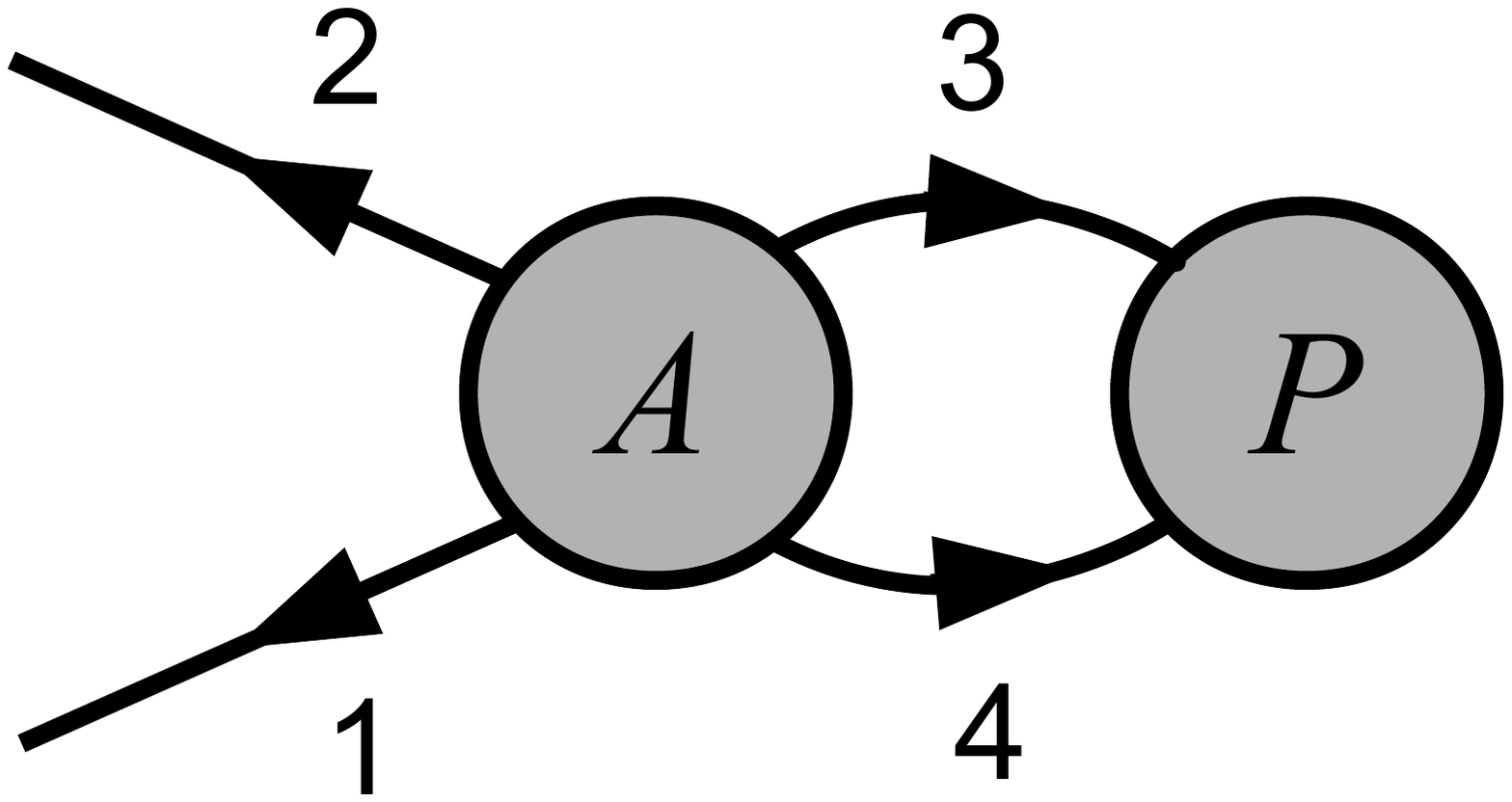}
\end{center}
\caption{\it In {\rm \cite{Zwiebel:2011bx}} it was shown that the harmonic
  action \eqref{harm} is recovered via the sewing together of a
  tree-level four-point superamplitude ${A}$ and a tree-level
  form factor ${P}$ corresponding to the particular two-site spin-chain
  state under consideration.}
\label{fig:1lff}
\end{figure}

A  direct  connection between the one-loop
nearest-neighbour part of the spin-chain dilatation operator and amplitudes, which will be very relevant for our investigation,  was found in 
\cite{Zwiebel:2011bx}  by Zwiebel, working off of an earlier
observation of Beisert.  In that paper the one-loop dilatation operator, expressed in the so-called
``harmonic action"  form \cite{Beisert:2003jj}, was related to the integration of a four-point superamplitude glued
to a tree-level form-factor with two external legs
over the two-particle phase space, see Figure \ref{fig:1lff}. 
In \cite{Wilhelm:2014qua}, this connection was explained  in terms of one-loop  form factors 
of generic operators.%
\footnote{See also \cite{ Koster:2014fva,Nandan:2014oga, Brandhuber:2014pta, Brandhuber:2015boa, Loebbert:2015ova,Frassek:2015rka} for related work connecting amplitudes, form factors and the dilatation operator.}
Specifically, it was shown there that the result of   \cite{Zwiebel:2011bx} is the
coefficient of the discontinuity of a bubble integral associated with this one-loop form factor, 
and captures the ultraviolet-divergent part of the calculation.

In the following we will use  Zwiebel's formula to show that the invariance of the amplitudes under the Yangian, and certain special properties of the Yangian of $PSU(2,2|4)$, lead precisely to the expected result \eqref{main}.

The plan of the paper is as follows. In section \ref{adv} we review basic facts about the
one-loop dilatation operator and its various realisations. Furthermore, we review the Dolan-Nappi-Witten \cite{Dolan:2003uh} proof
of \eqref{main},  which relies on a special set of eigenstates and motivate the calculation of the commutator 
$[Q, H]$. 
In section \ref{proof} we present a novel proof using ideas from amplitudes that does
not rely on any choice of a basis of states.

\section{Review and motivation}
\label{adv}

In this section we review some important facts about the dilatation operator and Yangian symmetry. We will then motivate the calculation of the commutator $[Q,H]$ performed in the next section using the representation of the dilatation operator in terms of amplitudes and form factors found in \cite{Zwiebel:2011bx}.  

\subsection{States and the spinor-helicity formalism}

We consider single-trace local operators in $\cN=4$ SYM of the form $\Tr (\Phi_1\cdots \Phi_L)(x)$, where the letters $\Phi$ are taken from the list $F^{\alpha \beta}, \ \psi^{\alpha ABC}, \ \phi^{[AB]}, \ \bar\psi^{\dot\alpha A}, \ \bar{F}^{\dot\alpha \dot\beta}$ (and symmetrised covariant derivatives acting on them), where  $A=1, \ldots , 4$ is a fundamental $SU(4)$ index.  

It is well known \cite{Gunaydin:1984fk} that the operators can be described in terms of excitations of 
two pairs  of bosonic oscillators  and one pair of fermionic oscillators, satisfying 
\beq
\label{comrel}
[a_\alpha, a^{\dagger \beta}] = \delta_\alpha^\beta\, , \quad 
[b_{\dot\alpha}, b^{\dagger {\dot\beta}}] = \delta_{\dot\alpha}^{\dot\beta}\, , \quad 
\{d_A, d^{\dagger B}\} = \delta_A^B\, , \quad
\alpha, \beta=1,2, \  \dot\alpha, \dot\beta=1,2, \ A=1, \ldots , 4,
\eeq
where the map to the letters introduced above  is 
\beq
\label{dic1}
\bar{F} \leftrightarrow b^\dagger b^\dagger\, , \quad \bar\psi \leftrightarrow b^\dagger d^\dagger\, , \quad \phi\leftrightarrow d^\dagger d^\dagger \, , 
\quad
\psi \leftrightarrow a^\dagger d^\dagger d^\dagger d^\dagger \, , \quad F \leftrightarrow a^\dagger a^\dagger d^\dagger d^\dagger d^\dagger d^\dagger 
\ , 
\eeq
while for derivatives   $D \leftrightarrow a^\dagger b^\dagger$. For instance, the Konishi operator 
$K = \epsilon_{ABCD} \phi^{AB} \phi^{CD}$ is represented as $ \epsilon_{ABCD} d^{\dagger A}_1 d^{\dagger B}_1d^{\dagger C}_2 d^{\dagger D}_2|0\rangle$.

The commutation relations \eqref{comrel} can then be realised in terms of spinor-helicity  variables, commonly used to describe amplitudes. The map in this case is 
\begin{equation}
\begin{split}
\label{dic2}
a^{\dagger \alpha} &\leftrightarrow \, \lambda^\alpha\, , \quad
b^{\dagger \dot\alpha} \leftrightarrow \,\tilde\lambda^{\dot\alpha} ,
\quad d^{\dagger A} \leftrightarrow \, \eta^A\, \\
a_{\alpha} &\leftrightarrow \,
{\partial \over \partial\lambda^\alpha}\, ,  \ \, b_{\dot\alpha}
\leftrightarrow \,\frac{\partial}{\partial \tilde\lambda^{\dot\alpha}} ,
\,\,\, d_{A} \leftrightarrow \, \frac{\partial}{\partial\eta^A}\, ,
\end{split}\end{equation} and, as usual in $\cN=4$ SYM, we combine the $\lambda$,
$\tilde\lambda$ and $\eta$ variables into a single object $\Lambda^a
:= \big( \lambda^\alpha, \tilde{\lambda}^{\dot\alpha}, \eta^A \big)$.
In this formalism, a state is simply a polynomial in the $\Lambda$'s
satisfying the physical state condition of vanishing central charge at
each spin-chain site, i.e.~it has a sensible translation back to the
letters $F^{\alpha \beta}, \ \psi^{\alpha ABC}, \ \phi^{[AB]},
\ \bar\psi^{\dot\alpha A}, \ \bar{F}^{\dot\alpha \dot\beta}$ (and
symmetrised covariant derivatives acting on them), and we denote it
as $P(\Lambda_1, \dots , \Lambda_L)$. Again, the Konishi operator is
represented in this language as $\epsilon_{ABCD}(\eta^A_1
\eta^B_1)(\eta^C_2 \eta^D_2)$. We also note that in
\cite{Wilhelm:2014qua} it was observed that $P(\Lambda_1, \dots ,
\Lambda_L)$ is nothing but the minimal form factor of the operator
represented by the state via the dictionaries \eqref{dic1} and
\eqref{dic2}.%
\footnote{The term ``minimal" form factor was introduced in
  \cite{Brandhuber:2014ica} to denote form factors where the state
  contains exactly as many particles as fields, i.e.~the number of
  fields is the minimal number required to have a non-zero result at tree level.}

\subsection{The complete one-loop dilatation operator}

At one loop and in the planar limit only two neighbouring fields
interact, and the one-loop dilatation operator $H$ is the sum of
densities $H_{i i+1}$, i.e.~$H = \sum_{i=1}^L H_{i i+1}$, where $L$ is
the number of fields in the operator (or sites in the spin chain, of
which $H$ is the Hamiltonian), and $H_{i i+1}$ acts only on fields at
position $i$ and $i+1$. The complete one-loop dilatation operator was
derived in \cite{Beisert:2003jj}, with the result \beq
\label{W}
H_{12} \ = \ \sum_{j=0}^\infty 2 h (j) \, \mathbb{P}_{12, j}\ .  \eeq
Here $h(j)$ is the $j^{\rm th}$ harmonic number and $\mathbb{P}_{12,
  j}$ projects onto a two-particle state with total spin $j$. The same
paper also introduced an alternative representation of the dilatation
operator in terms of the oscillators introduced in \eqref{comrel}
termed ``harmonic action". It is this representation which will be
particularly relevant for us, and specifically a rewriting of the
harmonic action in an integral form which was found in
\cite{Zwiebel:2007cpa}. Written in terms of spinor-helicity variables
the action is: 
\be\label{harm} H_{12} \, P\big({\Lambda}_1,
{\Lambda}_2\big) \ = \ - {1\over \pi} \int_0^{2 \pi}
\!d\phi\int_0^{\pi\over 2}\!d\theta\, \cot\theta \Big[e^{2 i \phi}
  P\big({\Lambda'}_1, {\Lambda'}_2\big) \, - \, P \big({\Lambda}_1,
  {\Lambda}_2\big)\Big] \ .  
\ee 
Here by $P(\Lambda_1, \Lambda_2)$ we
mean $P( \cdots , \Lambda_1, \Lambda_2 , \cdots)$ where the dots stand
for all other fields in the state represented by $P$ that are not
involved in the interaction. Moreover the $\Lambda^\prime$'s represent
``rotated" spinor-helicity variables defined as \be
\left( \begin{array}{c} \lambda'_{1} \\ \lambda_{2}'
\end{array} \right) := \mathcal{U}
\left( \begin{array}{c}\lambda_{1} \\ \lambda_{2}
\end{array} \right)\, , \quad 
\left( \begin{array}{c} {\tilde \lambda}_{1}' \\ {\tilde \lambda}_{2}'
\end{array} \right) := \mathcal{U}^\ast
\left( \begin{array}{c}{\tilde \lambda}_{1} \\ {\tilde \lambda}_{2}
\end{array} \right)\, ,
\left( \begin{array}{c} \eta_{1}' \\ \eta_{2}'
\end{array} \right) := \mathcal{U}^\ast
\left( \begin{array}{c}\eta_{1} \\ \eta_{2}
\end{array} \right)\, ,
\ee
with the matrix $\mathcal{U}$ given by  
\beq
\mathcal{U} := \left( \begin{array}{cc} \cos \theta & - e^{i \phi} \sin \theta \\ \sin \theta &  e^{i \phi} \cos \theta
\end{array} \right)\ . 
\eeq Note that while the state $P$ satisfies the central charge
condition, the rotated state in general violates this.  The integration
over $\phi$ in \eqref{harm} is precisely enforcing the condition that
the action of $H_{12}$ on $P$ returns a physical state.

\subsection{Connection to form factors}
As a final ingredient, we review an alternative form of \eqref{harm}  that was also discussed in 
 \cite{Zwiebel:2011bx}.%
 \footnote{We note that \cite{Zwiebel:2011bx} credits unpublished work of Beisert for pointing out the connection between the 
 rotating oscillator form of the harmonic action \eqref{harm} and \eqref{zzz} below.}
This representation for the action of the one-loop dilatation operator
on a state $| 1,2\rangle$ has the form%
\footnote{
\label{foot}Strictly speaking, this equation is only
  true up to a numerical factor which we leave out for aesthetic
  reasons, and think of  as being absorbed into the amplitude. This
  factor is related to the cut of a one-loop bubble integral and 
 its relation to the renormalisation constant of the operator \cite{Wilhelm:2014qua} and  will  cancel in our final result  \eqref{finres} and \eqref{finresbis}.}
\beq
\label{zzz}
H_{12} |1,2\rangle \ = \ \int\!d\Lambda \ A(1,2,3,4)\big[ P (-4, -3)  \,  - \,    r\,   P (1,2) \big]
\ , 
\eeq
where momentum conservation reads $p_1 + p_2 + p_3 + p_4 = 0$. $p_1$ and $p_2$ are the external legs, while $p_3$ and $p_4$ are integrated over with the appropriate two-particle phase-space  measure 
\beq
d\Lambda = \prod_{i=3}^4 d^2\lambda_i d^2\tilde{\lambda}_i d^4\eta_i
\ . 
\eeq
Note that
\be
A(1,2,3,4)= \frac{\delta^{(4)}(p)\,\delta^{(8)}(q)}{\la 12\ra\la 23\ra\la 34\ra\la 41\ra}, 
\ee
and the labels $1,\ldots, 4$ are a shorthand notation for $\Lambda_1, \ldots, \Lambda_4$. 
We have also defined the ratio 
\beq
\label{r}
r \ = \ \left({\langle 12\rangle \over \langle 34 \rangle }\right)^2
\ , 
\eeq
which allows us to write the two terms in \eqref{zzz} as integrated against the same tree-level amplitude, slightly departing from \cite{Zwiebel:2011bx} and \cite{Wilhelm:2014qua}. We find our presentation convenient  as it makes the infrared finiteness of \eqref{zzz} more manifest.

The relation between the two expressions for the dilatation
operator~\eqref{harm} and~\eqref{zzz} was shown
in~\cite{Zwiebel:2011bx}. After integrating out the
momentum conserving delta functions there are only two non-trivial
integrals left, over $\theta$ and $\phi$. The measures are then related by
\beq\label{measures}
d\Lambda\,  \big[  A(1,2,3,4)  \cdot r \big]   \rightarrow - {2\over 2 \pi}  \, d\phi\,  d\theta \,  \cot \theta
\ , 
\eeq
and we also have  $r \rightarrow e^{- 2 i \phi}$, $\Lambda_3\rightarrow
-\Lambda'_2$ and $\Lambda_4\rightarrow
-\Lambda'_1$. These replacements take us from~\eqref{zzz}
to~\eqref{harm}. 
As mentioned in  footnote  \ref{foot}, 
\eqref{measures} is strictly only true up to a multiplicative
numerical coefficient  which will
cancel in our final result.

Two observations are in order here. 

{\bf 1.} An important feature of \eqref{zzz} is that  it can be evaluated in four dimensions. The first term on the right-hand side of \eqref{zzz} has an infrared divergence which is cancelled by the second term. This can be understood  by observing that because of the four-point kinematics, the amplitude $A(1,2,3,4)$ develops a simple pole in the  forward-scattering limit 
\beq
\label{fsl}
p_4 \, = \, - p_1\, \qquad p_3 \, = \, - p_2 
\ , 
\eeq
which in turn generates infrared divergences in the first term of \eqref{zzz}.  
It is then clear that the second term in \eqref{zzz}  removes the  pole in the integration.%
\footnote{Similar considerations were made in \cite{Brandhuber:2009kh} in order to compute the  dual conformal anomaly of one-loop superamplitudes with arbitrary helicity.}

{\bf 2.} The fact that \eqref{zzz} provides a representation of the
complete one-loop dilatation operator of $\cN=4$ SYM may seem rather
mysterious thus far. A neat physical interpretation of this result was
found in  \cite{Wilhelm:2014qua}. In that paper it was observed that
the first term on the right-hand side of \eqref{zzz} is nothing but
the discontinuity (or two-particle cut) of a  one-loop minimal form
factor of a generic  operator. This one-loop form factor is
ultraviolet as well as infrared divergent, but the second term in
\eqref{zzz} removes  this infrared divergence, leaving only
ultraviolet divergences. At one loop, the latter are entirely captured
by a bubble integral, whose discontinuity is a finite numerical
constant. The coefficient of this discontinuity is  minus the one-loop dilatation operator, and this is precisely the right-hand side of \eqref{zzz}   \cite{Wilhelm:2014qua}.

\subsection{The Dolan-Nappi-Witten proof of the commutation relation}

The commutator $[Q,H]$  of the one-loop dilatation operator, $H$,  with a level-one
Yangian generator,%
\footnote{Note that  our definition of $Q_{12}$ is identical to that of \cite{Drummond:2009fd}, and differs from that of \cite{Dolan:2003uh} by a factor of $-1/2$, namely $Q_{12}^{\rm DNW} = (-1/2) Q_{12}^{\rm DHP}$. The  minus sign arises from having swapped the indices $B$ and $C$ in  \eqref{gener} compared to the corresponding definition in \cite{Dolan:2003uh}, while a  factor of $1/2$ is introduced in lowering an index of the structure constants in the definition of the Yangian generators in \cite{Dolan:2003uh}.} 
\beq
\label{gener}
Q^A := \sum_{i<j} Q_{ij}^A\, , \qquad Q_{ij}^A = f^A_{CB} J_i^B J_j^C\, , 
\eeq
where $J^A=\sum_i J_i^A$ are  level-zero (or superconformal) generators, 
was first examined in~\cite{Dolan:2003uh}. It was found to be given by a boundary term
\begin{align}
\label{sss}
  [Q^A, H]=2(J^A_1-J^A_L)\, , 
\end{align}
for a spin chain of length $L$.

The main ingredient in their proof of this was the two-body version of \eqref{sss}, namely
\begin{align}\label{eq:1}
  [ Q_{12}^A, H_{12} ] =2(J_1^A - J_2^A)
\end{align}
which they were then able to lift to the full $L$-site version.
We wish to give an alternate derivation of this formula in the next
section, but first, for comparison, we
remind readers of the original derivation of~\cite{Dolan:2003uh}. 

The derivation of~(\ref{eq:1}) in~\cite{Dolan:2003uh} relied on three
facts.

\begin{itemize}
\item[{\bf 1.}] It is possible to choose a basis for the two-body
  problem which simultaneously diagonalises the one-loop dilatation
  operator and the quadratic Casimir. That is, any 
  two-particle state  can be written as the sum of spin $j$ states
  $ |1,2\rangle=\sum_j |\lambda(j)\rangle$ where
  \begin{align}
    H_{12} |\lambda(j)\rangle=  2h(j) |\lambda(j)\rangle\,,\qquad
    J_{12}^2 |\lambda(j)\rangle= j(j+1) |\lambda(j)\rangle\ .
  \end{align}
Here  
$  J_{12}^2= (1/2) (J_1^A+J_2^A)(J_1^A+J_2^A)
$
is the quadratic Casimir operator. This is simply the tensor decomposition of two one-particle states into
irreducible representations upon which the dilatation operator acts diagonally.

\item[{\bf 2.}]  The level-one Yangian can be written as the commutator
  \begin{align}\label{eq:2}
     Q_{12}^A= -\frac{1}{2} [J_{12}^2 , J_1^A-J_2^A]\ .
  \end{align}
This can be checked straightforwardly.

\item[{\bf 3.}] The action of $J_1^A-  J_2^A$ on a spin $j$ state is a linear
  combination of a spin $j-1$ and a spin $j+1$ state,%
  \footnote{The proof of this can be found in \cite{Dolan:2003uh}.}
  \begin{align}
    (J_1^A-J_2^A)|\lambda(j)\rangle = |\chi^A(j-1)\rangle +
    |\rho^A(j+1)\rangle\ .
  \end{align}
\end{itemize}
The proof  proceeds very simply by first inserting \eqref{eq:2} into the
commutator $[Q_{12}, H_{12}]|\lambda(j)\rangle$ and using the above facts. 
One arrives at 
\begin{align}
  [Q_{12}^A, H_{12}]|\lambda(j)\rangle = 2j \Big[h(j) -
  h(j-1)\Big]|\chi^A(j-1)\rangle  + 2(j+1)\Big[h(j+1) -
  h(j)\Big]|\rho^A(j+1)\rangle\ .
\end{align}
Finally  using the numerical identity $h(j) -h(j-1)=1/j$ one finds
rather remarkably that
\begin{align}
  [Q_{12}^A, H_{12}]|\lambda(j)\rangle =2\left(|\chi^A(j-1)\rangle +
  |\rho^A(j+1)\rangle\right) = 2(J_1^A-J_2^A)|\lambda(j)\rangle\ .
\end{align}
Two comments are in order here. First, 
we note that while the proof relies heavily on choosing a specific
diagonal basis the final result is independent of any basis and is
purely an operator equation $ [Q_{12}^A, H_{12}]=2(J_1^A-J_2^A)$. We
wish to find a way to see this operator equation directly, and to make contact with the Yangian symmetry of amplitudes.  We will do
this in   section \ref{proof}.
Second, \eqref{eq:1} is a remarkable equation, in that the left-hand side is a one-loop quantity, while the right-hand side looks like  tree level. The key relation which allows for this is of course the identity $h(j) -h(j-1)=1/j$, and we wish to find a corresponding explanation from the amplitude point of view.

\subsection{Direct evaluation of the commutator $[Q, H]$ using \eqref{harm}}

In this section and in the next we would like to elucidate the power of the representation \eqref{zzz} of the dilatation operator over its  ``integrated" form   \eqref{harm} in evaluating the commutator
$[Q_{12},H_{12}]$. 
To this end we begin by acting on this latter representation with a level-one Yangian generator. Doing so we find, 
 \begin{eqnarray}
[Q_{12},H_{12}] P(1,2 ) \ = \ -\frac{1}{\pi} \int_0^{2 \pi}\!d\phi\int_0^{\pi\over 2}\!d\theta\,
\cot\theta \hspace{-0.3cm}& \Big[ e^{2 i \phi} Q_{12}(1,2) P(1',2') - Q_{12}(1,2) P(1,2)  \nonumber \\
& \hspace{-0.7cm} -e^{2 i \phi} (Q_{12} P)(1',2') + (Q_{12} P)(1,2)\Big], 
\end{eqnarray}
where the notation $Q_{12}(1,2)$ indicates that the operator acts on
the variables with labels $1,2$ while e.g. $(Q_{12} P)(1',2')$ means
that we act with $Q_{12}$ on the state $P$ and evaluate the result at
$1', 2'$. Importantly the second and fourth terms cancel each other and we are left with
\be
\label{kk}
[Q_{12},H_{12}] P(1,2 ) \ = \ -\frac{1}{\pi}\int_0^{2 \pi} \!d\phi\int_0^{\pi\over 2}\!d\theta\,
\cot\theta \,  e^{2 i \phi} \ \Big[ Q_{12}(1,2) P(1',2')  - (Q_{12} P)(1',2') \Big] \ . 
\ee
This integral is supposed to evaluate simply to 
\be
\label{myst}
[Q_{12},H_{12}] P(1,2 ) \ = \ 2 (p_1-p_2) P(1,2) \ , 
\ee
as we have  checked explicitly in a number of cases, however it is not obvious to see why this is true in general starting from \eqref{kk}. It is precisely this feature that we are going to demonstrate in the next section using the representation \eqref{zzz} provided by \cite{Zwiebel:2011bx}, and using the known action of Yangian generators on  tree-level scattering amplitudes.

\section{Proof of the commutation relations from amplitudes}
\label{proof}

We now come to the main part of this paper, where we evaluate the commutator $[Q, H]$ using the expression for $H$ in terms of amplitudes of  \cite{Zwiebel:2011bx} and the known action of Yangian generators on amplitudes   \cite{Drummond:2008vq, Drummond:2009fd}. In this way 
we both give a very simple proof of \eqref{myst}
 and at the same time further substantiate the connection between the
 spin chain and amplitude Yangians. 

\subsection{The commutator with the level-one Yangian generator $p^{(1)}$}
We wish to compute the commutator $[Q, H] |1,2\rangle$, where $|1,2\rangle$ is a two-particle state in the spin chain, and the $Q$ generators are defined in \eqref{gener}. 

As  discussed in \cite{Dolan:2003uh}, the  calculation of $[Q, H] |1,2\rangle$ boils down to that of the commutator 
$
[Q_{12}, H_{12}] |1,2\rangle$, 
which is what we address in this section.  Specifically, we will now discuss the case of $Q = p^{(1)}$, namely the generator corresponding to dual special conformal transformations $K$, and later consider the case $Q = q^{(1)}$, namely dual special conformal supersymmetry $S$.  
The commutator in question is  equal to
\beqa
\label{Qvar}
[Q_{12}, H_{12}] |1,2\rangle & = & Q_{12} \int\!d\Lambda \ A(1,2,3,4)\big[ P (-4, -3)   - r\,  P (1,2) \big] \nonumber \\ 
& - & 
\int\!d\Lambda \ A(1,2,3,4)\big[ Q_{-4, -3} P (-4, -3)   - r\,  Q_{12}   P (1,2)  \big] \ , 
\eeqa
where \cite{Drummond:2009fd}
\beq
\label{QQ}
Q_{ij} \ = \ \Big( m^{\, \gamma}_{j\, \, \alpha}\delta^{\dot\gamma}_{\, \dot\alpha} + 
\bar{m}^{\, \dot\gamma}_{j\, \, \dot\alpha}\delta^{\gamma}_{\, \alpha} - d_j \delta^{\gamma}_{\, \alpha} \delta^{\dot\gamma}_{\, \dot\alpha}\Big) p_{i\, \gamma \dot\gamma} \, + \, \bar{q}_{j \dot\alpha C} q^{C}_{ i \alpha} - ( i \leftrightarrow j ) \ . 
\eeq
The relevant generators are given by 
\beq
d_i \  = \ {1\over 2} \Big( \lambda_i^\alpha {\partial \over \partial \lambda_i^\alpha} + \tilde\lambda_i^{\dot\alpha} {\partial \over \partial \tilde\lambda_i^{\dot\alpha}} \Big) + 1\ , 
\eeq
and 
\beq
m_{\alpha \beta} = \lambda_{(\alpha} \partial_{\beta)}\, ,  \qquad \bar{m}_{\dot\alpha \dot\beta} = \tilde\lambda_{(\dot\alpha} \partial_{\dot\beta)}, \qquad   q^{A}_{\alpha} = \lambda_{\alpha} \eta^A \, ,  \qquad 
\bar{q}_{\dot\alpha A} = \tilde\lambda_{\dot\alpha} \partial_{A} \,   ,\qquad p_{\alpha \dot\alpha} = \lambda_\alpha \tilde\lambda_{\dot\alpha}
\ .
\eeq
We also note that $Q_{-4,-3}  = Q_{34}$.  Furthermore, in the second
line, $Q$ acts only on the form factor $P$, as required by the commutator. 

Before computing $[Q_{12}, H_{12}] |1,2\rangle $ for an arbitrary state, we find it instructive to discuss separately the case of a half-BPS operator. 

\subsection*{The commutator for a half-BPS state}

We consider the form factor representing the operator $\phi^{12}
\phi^{12}$, namely 
\beq
P^{\phi^{12} \phi^{12}}(1,2)  \ = \ \eta_{1}^1 \eta_{1}^2 \eta_{2}^1 \eta_{2}^2\ , 
\eeq
where lower indices denote the site, and upper indices the $R$-charge. 
The crucial fact about half-BPS operators is that 
\beq
 \int\!d\Lambda \ A(1,2,3,4) \, P (-4, -3)    \ = \      P (1,2)  \int\!d\Lambda \   A(1,2,3,4)    \cdot r 
 \ , 
 \eeq
as follows from the explicit calculation of \cite{Penante:2014sza}. Alternatively, this can be shown by  noticing that  
\beq
\eta_{3}^1 \eta_{3}^2 \eta_{4}^1 \eta_{4}^2 \ = \ r \, 
\eta_{1}^1 \eta_{1}^2 \eta_{2}^1 \eta_{2}^2 
\ , 
\eeq
as follows from supermomentum conservation $\sum_{i=1}^4 \lambda_i \eta_i = 0$.
As a consequence, the first line of \eqref{Qvar} vanishes when evaluated on a half-BPS state. 
We now evaluate the second line. Because this operator contains only scalars, it follows that  all terms  inside $Q_{ij}$ that contain spinor derivatives vanish. Because the operator is half BPS, it also follows that the $\bar{q} q$ term  in \eqref{QQ} annihilates the operator. The only surviving contribution is  that arising from the constant part in the dilatation operator inside  \eqref{QQ}. We then find that 
\beqa
Q_{12} P^{\phi^{12} \phi^{12}}(1,2) &=& - (p_1 - p_2) P^{\phi^{12} \phi^{12}}(1,2) 
\ , \nonumber \\ \cr
Q_{-4-3} P^{\phi^{12} \phi^{12}}(-4,-3) &=& - (-p_4 + p_3) P^{\phi^{12} \phi^{12}}(-4,-3)\nonumber 
\\ &=& -   r (-p_4 + p_3)
P^{\phi^{12} \phi^{12}}(1,2)  \ , 
\eeqa
thus
\beq
\label{bpsvar}
[Q_{12}, H_{12}] |\phi^{12} \phi^{12} \rangle =  P^{\phi^{12} \phi^{12}} (1,2)
\int\!d\Lambda \ A(1,2,3,4) \cdot r \, \big[ p_3 - p_4 - (p_1 - p_2) \big]  \ .  
\eeq
Again, note that \eqref{bpsvar} is a finite integral, as the region
responsible for infrared divergences, $p_4=-p_1$, $p_3=-p_2$,
explicitly makes the term in the square brackets vanish.
We can now evaluate the remaining integral using the parameterisation introduced in \cite{Zwiebel:2011bx}.
All variables except $\theta$ and $\phi$ can be integrated trivially using delta functions, and one is left with the following effective parameterisation for the loop momenta, 
\beqa
\label{Zw}
\lambda_3 & = & \lambda_1 \sin \theta \, + \, e^{i \phi}  \lambda_2 \cos \theta\ , \qquad 
\tilde\lambda_3 \ = \  - ( \tilde\lambda_1 \sin \theta \, + \, e^{-i \phi}  \tilde\lambda_2 \cos \theta)\ ,
\nonumber \\
\lambda_4 & = & \lambda_1 \cos \theta \, - \, e^{i \phi}  \lambda_2 \sin \theta\ , \qquad 
\tilde\lambda_4 \ = \  - ( \tilde\lambda_1 \cos \theta\, - \, e^{-i \phi}  \tilde\lambda_2 \sin \theta)\ ,
\,  
\eeqa
We then find 
\beq
\label{p3-p4}
p_3 - p_4 - (p_1 - p_2) \ = \ 2 \Big[  \sin^2 \theta \,(p_2 - p_1) - \cos \theta\sin \theta \, ( \lambda_1 \tilde\lambda_2 e^{- i \phi} + \lambda_2 \tilde\lambda_1e^{ i \phi} ) \Big]\ . 
\eeq
As shown in \cite{Zwiebel:2011bx}, the integration measure $d\Lambda\,  A(1,2,3,4)$ in \eqref{bpsvar} becomes, after integrating out all delta functions,%
\footnote{The normalisation in \eqref{result} is such that \eqref{zzz} agrees with \eqref{W}. It is at this point that the numerical factor mentioned in footnote \ref{foot} cancels out. 
We also remind the reader  that in the parameterisation \eqref{Zw} one simply has $r = e^{- 2 i \phi}$.
} 
\beq\label{result}
d\Lambda\,  \big[  A(1,2,3,4)  \cdot r \big] \rightarrow  - {2\over 2 \pi}  \, d\phi\,  d\theta \,  \cot \theta
\ , 
\eeq
where $\theta \in (0, \pi/2)$ and $\phi \in (0, 2 \pi)$. Using \eqref{p3-p4} and \eqref{result} one then finds 
\beq
\label{finres}
\int\!d\Lambda \ A(1,2,3,4)\cdot r \, \big[ p_3 - p_4 - (p_1 - p_2) \big]  \ = \ 2 \, (p_1 - p_2)
\ , 
\eeq
where terms proportional to $e^{\pm i \phi}$ in  \eqref{p3-p4} trivially integrate to zero. In conclusion, we find 
\beq
[Q_{12}, H_{12}] |\phi^{12} \phi^{12} \rangle =   2 \, (p_1 - p_2) |\phi^{12} \phi^{12} \rangle \ , 
\eeq
in agreement with \cite{Dolan:2003uh}.

\subsection*{The commutator for generic states}

After this detour we go back to our proof.  First, we observe  that we can rewrite \eqref{Qvar} as 
\beqa
\label{Qvar2}
[Q_{12}, H_{12}] |1,2\rangle &= &  \int\!d\Lambda \ \big[ (Q_{12} + Q_{34}) A(1,2,3,4)\big] \big[ P (-4, -3)   - r  P (1,2) \big] \nonumber \\ 
&-& 
 \int\!d\Lambda \ \big[ (Q_{34} - ( p_3 - p_4)  \big]  \Big[ A(1,2,3,4)\big[ P (-4, -3)   -r   P (1,2) \big] \Big] 
\nonumber \\ 
&-& 
P (1,2) \,  \int\!d\Lambda \ \Big[\big(\hat Q_{12}+\hat Q_{34}\big)\, r\Big]  \,  A(1,2,3,4)  
 \nonumber \\ 
&-& 
P (1,2) \,  \int\!d\Lambda \ \big(p_1 -  p_2-  p_3 + p_4  \big)  \,  A(1,2,3,4) \cdot r  \ . 
\eeqa
In going from \eqref{Qvar} to \eqref{Qvar2}  we have performed an integration by parts, taking special care of the multiplicative part of $Q_{ij}$, obtained from taking the constant piece inside the dilatation operator.  We have defined $\hat Q_{ij}$ to be the differential part of $Q_{ij}$, that is $\hat Q_{ij}:=Q_{ij}+p_i-p_j$.

We will now show that the following statements concerning \eqref{Qvar2} are true: 
\begin{itemize}
\item[{\bf 1.}] The first line vanishes due to two reasons: first, $\sum_{i<j}Q_{ij}$ is the dual conformal generator $K$ (up to a linear combination of level-zero generators, which annihilate the amplitude), which is a symmetry of the amplitudes; and second,  the  nature of the supergroup $PSU(2,2|4)$, and specifically the  vanishing  of its dual Coxeter number. 

\item[{\bf 2.}] The second line is a total derivative and integrates to zero. 

\item[{\bf 3.}] We show that $(\hat Q_{12}+\hat Q_{34})\, r =0$ and hence  the third line vanishes.

\item[{\bf 4.}] The last line is the only non-zero contribution and
  provides the expected answer for the commutator. This is shown
  explicitly below.

\end{itemize}

\noindent
{\bf 1.} We rewrite $Q_{12} + Q_{34} = \sum_{i< j} Q_{ij}  - (Q_{13} + Q_{14} +Q_{23} + Q_{24})$. We then observe that 
$ \sum_{i< j} Q_{ij}$ is precisely a Yangian generator, which annihilates the tree amplitude \cite{Drummond:2009fd}.
We can then recast the second term as\footnote{We note the
  similarity between the right-hand side of \eqref{dcn} and Eq.~(3) of
  \cite{Beisert:2011pn}.}  
\beq
\label{dcn}
(Q_{13} + Q_{14} +Q_{23} + Q_{24})^A = f^{A}_{CB} (J_1 + J_2)^B (J_3 + J_4)^C\,  = \, f^{A}_{CB} (J_1 + J_2)^B J^C \,  -\,  {1\over 2} f^{A}_{CB} f^{BC}_{D} (J_1 + J_2)^D \ , 
\eeq
where $J:= J_1 + \cdots +J_4$. 
The last term in \eqref{dcn} is proportional to the dual Coxeter number of $PSU(2,2|4)$ and hence vanishes. 
The penultimate term in \eqref{dcn} contains a level-zero generator $J^C$, which annihilates the amplitude. Thus \beq
(Q_{13} + Q_{14} +Q_{23} + Q_{24})\, A(1,2,3,4) = 0
\ . 
\eeq 
There is another way to appreciate this. Indeed, the fact that $Q_{13} + Q_{14}
+Q_{23} + Q_{24}$ annihilates the amplitude is due to the fact that Yangian symmetry is compatible with the 
cyclicity of amplitudes. In more detail, 
\beq
\sum_{1\leq i <j\leq 4} Q_{ij} - \sum_{3\leq i <j\leq 6} Q_{ij} \ = \ 2 (Q_{13} + Q_{14} +Q_{23} + Q_{24})
\ , 
\eeq
where we identify particle $i$ with $i+4$.  The two expressions $\sum_{1\leq i <j\leq 4} Q_{ij}$ and $\sum_{3\leq i <j\leq 6} Q_{ij}$ provide two representations of the level-one Yangian generator differing by a shift by two units of the particle labels. It is known from the work of  \cite{Drummond:2009fd} that the Yangian is consistent with the cyclicity of the scattering amplitudes, hence both expressions annihilate the tree amplitude.

\noindent
{\bf 2.} 
We consider the  second term in \eqref{Qvar2}, which contains the combination $Q_{34} - ( p_3 - p_4)$,  and show that it can be rewritten as a total derivative. 
Looking at the expression for $Q_{ij}$ in \eqref{QQ}, we note that the terms involving $m$, $\bar{m}$ and $\bar{q} q$ are total derivatives. We only need to focus on the term involving the tree-level dilatation operator $d$. To this end we note that relevant term is $-d_4 p_3 + d_3 p_4 - p_3 + p_4 = - (d_4 + 1)p_3 + (d_3 + 1)p_4$. 
We can then write its action on a function $f$ as a total derivative, 
\beq
\label{total}
(1 + d_i)f = \Big[ 2 + {1\over 2} \Big( \lambda_i^\alpha {\partial \over \partial \lambda_i^\alpha} + \tilde\lambda_i^{\dot\alpha} {\partial \over \partial \tilde\lambda_i^{\dot\alpha}} \Big)\Big] f   \ = \ 
{1\over 2} \Big[ {\partial \over \partial \lambda_i^\alpha}     ( \lambda_i^\alpha \, f) + {\partial \over \partial \tilde\lambda_i^{\dot\alpha}} ( \tilde\lambda_i^{\dot\alpha} f)\Big]
\ . 
\eeq
The second line in \eqref{Qvar2} is then a boundary term which vanishes. Note that the integration can be carried out in four dimensions since the integral is finite.

\noindent{\bf 3.} 
A short calculation shows that the stronger statements   
\beq
\hat{Q}_{12}\,  r \, = \, \hat{Q}_{34} \, r = 0 \ , 
\eeq
are true. Since $r = e^{-2 i \phi}$ and the integration over $\phi$ imposes the vanishing of the central charge on the physical states,  this condition should be equivalent to the fact that the central charge commutes with all generators of the algebra and hence also with $\hat{Q}$. 

\noindent{\bf 4.}
Finally the last term is the only one that contributes to the commutator. It  was in fact calculated earlier in \eqref{finres}, and crucially, it is proportional to the tree-level form factor $P(1,2)$. Using this result, we get  
\beq
\label{finresbis}
- P (1,2) \,  \int\!d\Lambda \ \big(p_1 -  p_2-  p_3 + p_4  \big)  \,  A(1,2,3,4)  \cdot r \ = \ 2 \, (p_1 - p_2)\, P (1,2)
\ . 
\eeq
A final comment is in order before concluding this section. One should exercise some caution in the manipulations above, in particular in setting $K_{\alpha \dot\alpha} A = 0$.  In fact, 
$K_{\alpha \dot\alpha} A $ contains a yet unnoticed holomorphic anomaly \cite{Cachazo:2004by}
arising only in   four-point kinematics. The key fact to notice is that \cite{Korchemsky:2009hm}
\beq
\label{holoanom}
K_{\alpha \dot\alpha} {1\over \langle i\,  i\!+\!1\rangle} \ = \ 2 \pi \, \delta( \langle i \, i\!+\!1\rangle) \delta( [ i \, i\!+\!1])\, [i \, i\!+\!1]\, (p_i + p_{i+1})_{\alpha \dot\alpha}
 \ . 
 \eeq
 The right-hand side of \eqref{holoanom} vanishes, unless the $[i \, i\!+\!1]$ factor is compensated by a corresponding pole, which indeed occurs  in a four-point amplitude $A(1,2,3,4)$, when, for instance, the vanishing of $\langle 23\rangle$ implies the vanishing of  $\langle 41\rangle$. 
Such a holomorphic anomaly could affect  the first and second line of \eqref{Qvar2}. However, thanks to the presence of the combination $P(-4, -3) - r P(1,2)$, which precisely vanishes on the support of the delta function, i.e. the forward-scattering kinematic configuration, these holomorphic anomalies cancel out. 
 
In conclusion, we have demonstrated that 
\beq
[Q_{12}, H_{12}] |s\rangle  \ = \ 2 (p_1 - p_2)  |s\rangle
\ . 
\eeq
This is the main result of the paper.
In the remaining subsection we work out additional examples of commutators with level-one and level-zero generators.

\subsection{Additional commutators}

In principle it is not necessary to check commutators with other level-one generators, given
the invariance of $H$ under the standard superconformal group. Nevertheless, we give here
the proof for the case of $q^{(1)}$, which is very similar to that for $p^{(1)}$. Specifically, 
\eqref{Qvar2} still holds with $Q = q^{(1)}$ and each momentum $p_i$ replaced by the corresponding supermomentum $q_i$. In order to convince ourselves of this fact, we recall that 
\beq
Q_{ij} \, := \, m_{j \alpha}^\gamma  q_{i\gamma}^A - {1\over 2} (d_j + c_j)  q_{i\alpha}^A + p_{j \alpha}^{\dot\beta} \bar{s}^A_{i \dot\beta} + q_{j \alpha}^B r^A_{i B} \,  - \, (i \leftrightarrow j ) 
\ . 
\eeq
The only difference occurs in point {\bf 2.} of the previous discussion. In particular, \eqref{total} is replaced by 
\beq
\Big[1 + {1\over 2} ( d_i + c_i) \Big] f = 
\Big[ 2 + {1\over 2}  \lambda_i^\alpha {\partial \over \partial \lambda_i^\alpha} -{1\over 4} \eta^A {\partial \over \partial \eta^A}\Big] f   \ = \ 
{1\over 2} {\partial \over \partial \lambda_i^\alpha}     ( \lambda_i^\alpha \, f) + {1\over 4} {\partial \over \partial \eta^A} ( \eta^A f)
\ . 
\eeq
We also comment that, as in the previous case, the  derivative part of the operators $Q_{12}$ and $Q_{34}$ commute with $r$ defined in \eqref{r}. 


Our main result relies crucially on integration by parts involving  the level-zero dilatation
operator and we would like to demonstrate that its commutation relation with $H_{12}$ 
indeed vanishes in this amplitude-based approach. Note that invariance under  
Lorentz transformations was explicitly checked in  \cite{Zwiebel:2011bx}, but the case
of dilatations is slightly more subtle.
This calculation can be performed efficiently by noticing that replacing $Q$ with $d$ in \eqref{Qvar2} is equivalent to performing the following replacement in that equation,  
\beq
p_1 \to -1\, , \qquad p_2 \to 1\,, \qquad p_3 \to -1\, , \qquad p_4 \to 1 \ . 
\eeq
The  second line then becomes $d_1+d_2+2$, which crucially is equal to a total derivative,  
\beq
\Big(2 +   d_1 + d_2 \Big) f  
  \ = \ 
\sum_{i=1}^2{1\over 2}  \Big[ {\partial \over \partial \lambda_i^\alpha}     ( \lambda_i^\alpha \, f) + {1\over 2} {\partial \over \partial \tilde\lambda_i^{\dot\alpha}}     ( \tilde\lambda_i^{\dot\alpha} \, f) \Big]
\ .
\eeq
The remaining lines in \eqref{Qvar2} are then easily seen to vanish as well.

\section*{Acknowledgements}

It is a pleasure  to thank  Matthias Staudacher and Matthias Wilhelm for bringing the paper \cite{Zwiebel:2011bx} to our attention, and in particular Matthias Staudacher for an inspiring discussion on Beisert's harmonic action. We would also like to thank Florian Loebbert and Jan Plefka for a useful discussion on \cite{Dolan:2003uh}, and Niklas Beisert, Rouven Frassek, Martyna Kostacinska and Brenda Penante   for  related conversations. 
GT would like to thank the Department of Mathematical Sciences and Grey College at Durham University for their warm hospitality through a Grey Fellowship. 
The work of AB, GT and DY was supported by the Science and Technology Facilities Council Consolidated Grant ST/L000415/1  
{\it ``String theory, gauge theory \& duality"}, while that of  PH was supported by the the Science and Technology Facilities Council Consolidated Grant 
ST/L000407/1 {\it ``Particles, fields and spacetime"}.

\bibliographystyle{utphys}
\bibliography{yangian}
\end{document}